\documentclass[11pt]{article}

\usepackage{epsfig}
\usepackage{a4}

\newif\iffigframe

\newif\ifbigfig

\newlength{\dinwidth}
\newlength{\dinmargin}
\newcommand{\figbox}[3]{\hbox to#1\bgroup
  \dimen0=1bp \dimen1=#1\relax
  \def\a##1 ##2 ##3 ##4 ##5\\{\if!##1!\a##2 ##3 ##4 ##5 .\\\else
    \dimen3=##3\dimen0 \advance\dimen3 -##1\dimen0
    \dimen4=##4\dimen0 \advance\dimen4 -##2\dimen0
    \dimen5=\dimen4 \divide\dimen5 \dimen3
    \dimen2=\dimen1 \multiply\dimen2 \dimen5
    \multiply\dimen5 \dimen3 \advance\dimen4 -\dimen5
    \dimen5=\dimen1
    \loop \advance\dimen4 \dimen4 \divide\dimen5 2
    \ifnum\dimen5>0 \ifnum\dimen4<\dimen3 \else
      \advance\dimen4 -\dimen3 \advance\dimen2 \dimen5 \fi
    \repeat
    \dimen5=10\dimen1 \divide\dimen5 \dimen0
    \includegraphics{#3.eps}%
    \iffigframe \vrule\hss \else \hfil \fi
    \vbox to\dimen2\bgroup
      \iffigframe \hrule width\dimen1\vss \hrule \else \vfil \fi
      \egroup
    \iffigframe \vrule\hss \fi
    \egroup\fi}%
  \a#2 . . . .\\}

\newcounter{subequation}[equation]

\makeatletter
\expandafter\let\expandafter\reset@font\csname reset@font\endcsname
\newenvironment{subeqnarray}
  {\arraycolsep1pt
    \def\@eqnnum\stepcounter##1{\stepcounter{subequation}{\reset@font\rm
      (\theequation\alph{subequation})}}\eqnarray}%
  {\endeqnarray\stepcounter{equation}}
\makeatother

\newcounter{statement}
\newenvironment{statement}[4]
  {\par\refstepcounter{statement}
    \noindent#1#2 \arabic{statement} #4\unskip: #3}{\par\vspace{2mm}}
\newenvironment{Thm}
  {\begin{statement}{\bf}{Theorem}{\sl}}{\end{statement}}
\newenvironment{Prop}
  {\begin{statement}{\bf}{Proposition}{\sl}}{\end{statement}}
\newenvironment{Lemma}
  {\begin{statement}{\bf}{Lemma}{\sl}}{\end{statement}}
\newenvironment{statement*}[4]
  {\par\noindent#1#2 #4\unskip: #3}{\par\vspace{2mm}}
\newenvironment{Coro}
  {\begin{statement*}{\bf}{Corollary}{\sl}}{\end{statement*}}
\newenvironment{Prf}
  {\begin{statement*}{\sl}{Proof}{\rm}}{\end{statement*}}

\begin{document}

\hbox to\hsize{%
  \vbox{%
        }\hfil
  \vbox{%
        \hbox{MPP-2004-49}%
        \hbox{\today}%
        }}

\vspace{1cm}
\begin{center}
\LARGE\bf
Static, spherically symmetric solutions
of Yang-Mills-Dilaton theory.
\vskip5mm
\large Dieter Maison\\
\vspace{3mm}
\small\sl
Max-Planck-Institut f\"ur Physik\\
--- Werner Heisenberg Institut ---\\
F\"ohringer Ring 6\\
D-80805 Munich, Germany\\
\end{center}
\vspace{10mm}
\begingroup \addtolength{\leftskip}{1cm} \addtolength{\rightskip}{1cm}
\subsection*{Abstract}
Static, spherically symmetric solutions of the Yang-Mills-Dilaton theory are 
studied. It is shown that these solutions fall into three different classes.
The generic solutions are singular. Besides there is a discrete set of
globally regular solutions further distinguished by the number of nodes of
their Yang-Mills potential. The third class consists of oscillating solutions
playing the role of limits of regular solutions, when the number of nodes
tends to infinity. We show that all three sets of solutions are non-empty.
Furthermore we give asymptotic formulae for the parameters of regular
solutions and confront them with numerical results.   
\endgroup
\vspace{1cm}

\section{Introduction}\label{chapint}
The dilaton may be considered as a kind of scalar graviton sharing with it a
universal coupling to matter. From this point of view it may be not too
surprising that the static, spherically symmetric solutions of the
Yang-Mills-Dilaton (YMD) theory share many properties with their
Einstein-Yang-Mills (EYM) relatives.
In fact, numerical studies \cite{LM,Bizon} have revealed a great similarity
between a family of `gravitational sphalerons' -- the Bartnik-McKinnon (BK)
solutions \cite{Bart} -- and a corresponding family of dilaton solutions.
Also an existence proof of these solutions running exactly
along the lines of the one for the BK solutions \cite{Smoller} could be
given \cite{Hastings}. 

Introducing a `stringy' radial variable rescaled with a dilaton factor
the field equations of the YMD theory resemble very much those of the EYM
theory. At first sight the only difference is the larger number of
gravitational degrees of freedom, which is however upset by the radial
diffeomorphism constraint of the EYM theory. Thus there is practically no
difference between the two theories concerning the number of degrees of
freedom. In \cite{BFM} a classification of all static, spherically
symmetric solutions of the EYM theory with a regular origin was given;
our aim is to prove a corresponding classification for the YMD theory.
Naively one could expect that this should be an easier task for the YMD
theory due to its simpler structure. However, life is not so simple and,
although ultimately the result is the same, the proof for the YMD theory 
appears to be more difficult. The main reason is that in the EYM theory the
`area variable' $r$ can have at most one maximum, its counterpart in YMD
theory can oscillate.
Apart from this subtlety things in the EYM and YMD theory turn out to be
largely the same. In both cases there are three different types of solutions
with a regular origin. The generic one develops a singularity of the
gravitational resp.\ dilaton field for a finite value of the corresponding
autonomous radial variable. The second type is a countably infinite family
of globally regular solutions differing by the number $n$ of nodes of the YM
potential $W$. Finally there is an oscillating limiting solution for
$n\to\infty$.
Based on this classification an existence proof for the three different
types of solutions for the EYM theory was given \cite{BFM}, 
which can be easily adapted to the YMD theory.
Futhermore the asymptotic scaling law of the parameters of the BK solutions
for large $n$ derived in \cite{BFM} can be straightforwardly translated to
the YMD theory.

For the {\sl SU(2)\/} Yang-Mills field $W_\mu^a$ we use the
standard minimal spherically symmetric (purely `magnetic') ansatz
\begin{equation}\label{Ans}
W_\mu^a T_a dx^\mu=
  W(R) (T_1 d\theta+T_2\sin\theta d\varphi) + T_3 \cos\theta
d\varphi\;,
\end{equation}
where $T_a$ denote the generators of {\sl SU(2)\/} and $R$ the radial
coordinate.

The action of YMD theory has the form
\begin{equation} \label{eymlaction}
S_{\rm YMD}=\int\Bigl(\frac{1}{2}\partial_\mu\phi\partial^\mu\phi
 -{1\over 4 g^2}e^{2\kappa\phi}F^{a}_{\mu\nu}F^{a \mu\nu}
\Bigr)d^4x \; ,
\end{equation}
where $g$ and $\kappa$ are the gauge resp.\ dilaton coupling.
Inserting the ansatz Eq.~(\ref{Ans}) into the action and making use of the
spherical symmetry we obtain the reduced YMD action
\begin{equation}\label{action}
S=\int dR \Biggl({R^2\phi'^2\over 2}+\frac{1}{g^2}e^{2\kappa\phi}\Bigl(
             W'^2+{(1-W^2)^2\over2R^2}\Bigr)\Biggr)\;.
\end{equation}
The dependence on $g$ and $\kappa$ can be removed by the rescaling
$\phi\to\phi/\kappa$, $R\to R\kappa/g$ and $S\to Sg\kappa$.

The resulting Euler-Lagrange equations are
\begin{subeqnarray}\label{feq}
   (R^2\phi')'&=&2e^{2\phi}\Bigl(W'^2+\frac{(1-W^2)^2}{2R^2}\Bigr)\;,\\
   W''&=&{W(W^2-1)\over R^2}-2\phi'W'\;.
\end{subeqnarray}
Using $\tau=\ln R$ as a coordinate and introducing
\begin{equation}\label{first}
r\equiv Re^{-\phi}, \quad N\equiv 1-R\phi',\quad U\equiv e^\phi W'
\quad T\equiv \frac{W^2-1}{r}\;,
\end{equation}
we obtain the autonomous first order system of Riccati type 
(the dot denoting a $\tau$ derivative)
\begin{subeqnarray}\label{taueq}
  \dot r&=&rN\;,\\
  \dot W&=&rU\;,\\
  \dot N &=&1-N-2U^2-T^2\;,\\
  \dot U&=&WT+(N-1)U\;,\\
  \dot T&=&2WU-NT\;
\end{subeqnarray}
supplemented by the constraint  $W^2-rT=1$.
These equations have a great similarity to the Eqs.~(50) of \cite{BFM}
for the Einstein-YM theory.
Clearly there is no analogue of the diffeomorphism constraint here. 

There is a kind of $\tau$-dependent `energy'
$E=2\dot W^2-(W^2-1)^2$ obeying 
\begin{equation}\label{energy}
\dot E=4(2N-1)\dot W^2\;,
\end{equation}
which will be useful as a `Lyapunov Function'. In addition we
introduce some other useful auxiliary quantities $e$, $f$ and $g$ defined
as
\begin{equation}\label{aux}
e\equiv \frac{E}{r^2}=2U^2-T^2\;,\quad f\equiv (1-N)^2+e\quad 
{\rm and}\quad g\equiv 1-N-f\;,
\end{equation}
obeying the equations
\begin{subeqnarray}\label{auxeq}
  \dot e&=&4(2N-1)U^2-2Ne\;,\\
  \dot f&=&-2f+4U^2\;,\\
  \dot g&=&-g+(1-N)^2\;.\\
\end{subeqnarray}
In general it does not seem possible to solve Eqs.~(\ref{taueq}) in closed
form, yet there are some simple exeptions.
One is the trivial vacuum solution $W^2=1$, $\phi=$const.; besides 
there is an
analogue of the extremal magnetically charged Reissner-Nord\-str{\o}m (RN) 
solution of the EYM theory
\begin{equation}\label{eRN}
W\equiv0 \quad N=1-\frac{1}{r}\quad \phi=-\ln(1+ce^{-\tau})\;.
\end{equation}
\section{Singular Points}\label{chapsing}

Before trying to explore the global behaviour of solutions
with a regular origin, it is important to know
the singular points of the system (\ref{taueq}). Actually, there
are two types of singular points, those attained for finite $\tau$
and the fixed points for $\tau\to\pm\infty$. The first type of 
singularity occurs, if  the r.h.s.\ of Eqs.~(\ref{taueq}) blows up
at some finite value of $\tau$. 
As we shall prove later, the only such possibility is that
$N\to -\infty$ and $r\to 0$. 
Due to the simple Riccati form of Eqs.~(\ref{taueq}) it is easy to find all 
their fixed points.
There are first the f.p.s $r=0,\infty$ and $N=1,W=\pm1,U=T=0$. 
Furthermore there is the f.p.\ $N=0,r=T=1,W=U=0$ like for the EYM theory.
All these f.p.s are of hyperbolic type and thus    
the application of the theory of dynamical systems
\cite{Codd,Arnold} provides theorems on local
existence and the asymptotic behaviour near the singular points.

Introducing suitable auxiliary variables it is also possible to 
treat the singularity with $N\to -\infty$ (and $r\to 0$) as a fixed point.
For that purpose it
turns out to be convenient to use $r$ as the independent variable 
and two auxiliary dependent variables
\begin{equation}\label{Rzero}
\kappa\equiv r(1-N), \quad \lambda\equiv WT+(N-1)U 
\end{equation}
obeying the equations
\begin{subeqnarray}\label{Req}
  \frac{d}{dr}W&=&-\frac{rU}{\kappa-r}\;,\\
  \frac{d}{dr}U&=&-\frac{\lambda}{\kappa-r}\;,\\
  \frac{d}{dr}\kappa&=&\frac{(f-4U^2)r}{\kappa-r}\;,\\
  \frac{d}{dr}\lambda&=&\frac{(1-3W^2+4U^2-f)U+\lambda}{\kappa-r}\;,\\
  \frac{d}{dr}f&=&\frac{2f-4U^2}{\kappa-r}\;.\\
\end{subeqnarray}
Applying Prop.~4 of \cite{BFM} assuming $\kappa\neq0$ we obtain
finite limits $W_0$, $U_0$, $\kappa_0$, $\lambda_0$, $\mu_0$ and $f_0$ 
for the dependent variables. 
The finiteness of $f$ implies $\kappa_0=|W_0^2-1|$
and thus we have to require $W_0^2\neq1$. 
From the behaviour $N\sim 1/r$ it follows that $\tau$ resp.\ $r$
stay finite as $r\to 0$ and hence $\phi\to\infty$. The original
dependent variables behave as
\begin{subeqnarray}\label{sing}
  W&=&W_0+\frac{W_0}{2(1-W_0^2)}r^2+W_3r^3+O(r^4)\;,\\
  U&=&\pm W_0-\Bigl(3|W_0^2-1|W_3+\frac{W_0}{W_0^2-1}\Bigr)r+O(r^3))\;,\\
  N&=&-\frac{|W_0^2-1|}{r}+1+N_1r+O(r^2)\;.
\end{subeqnarray}
with arbitrary parameters $W_0,W_3$ and $N_1$.
Actually $r=0$ is a regular point of Eqs.~(\ref{Req}) as long as 
$\kappa>0$ and thus this type of singular
behaviour with $N\to -\infty$ is of a generic type.

The case $W_0^2=1$ requires special treatment. Unlike the EYM
theory there are no solutions with a `Schwarzschild type` singularity
$W_0=\pm1$, $\kappa\sim |W_0^2-1|/r+N_0$ with $N_0\neq 1$
and solutions with
$W_0=\pm1$ automatically have a regular origin with $U=0$ and $N=1$.
The f.p.\ treatment of this case proceeds like for the
Bartnik-McKinnon solutions given in \cite{BFM} and we don't repeat it here.
The linearization at the f.p.\ gives the solutions
\begin{subeqnarray}\label{wsing}
  W&=&\pm(1-be^{2\tau}+ce^{-\tau})\;,\\
  N&=&1-4b^2e^{2\phi_0}e^{2\tau}\;\\
  \phi&=&\phi_0+2b^2e^{2\phi_0}e^{2\tau}+de^{-\tau}\;,
\end{subeqnarray}
where $b$, $c$ , $\phi_0$ and $d$ are free parameters.
For regular solutions, i.e.\ those running into the f.p.\ for $\tau\to
-\infty$
the coefficients $c$ and $d$ have to vanish.
Without restriction one may choose the $+$ sign for $W$ and
$\phi_0=0$; then one gets $W=1-br^2+c/r+O(r^4)$ etc.

Similarly there is the fixed point $W=\pm 1$, $U=0$ and $N=1$ with
$r\to\infty$ corresponding to asymptotically regular solutions.
The asymptotic behaviour close to the f.p.\ is again dominated by the linear
approximation
\begin{subeqnarray}\label{rinf}
  W&=&\pm(1+ce^{-\tau}+be^{2\tau})\;,\\
  N&=&1+de^{-\tau}\;,\\
  \phi&=&\phi_\infty+de^{-\tau}\;.
\end{subeqnarray}
Regular solutions require now $b=0$ and normalizing $\phi$ to
$\phi_\infty=0$ one obtains $W=\pm(1+c/r+br^2+O(r^{-2}))$ etc.

There is one more fixed point analogous to the
`RN fixed point' of the EYM theory given by $W=U=N=0$, $r=1$ and hence
$\phi\to\infty$. For simplicity we call it again the RN fixed point.
Putting $\bar r=r-1$ the linearization at this
f.p.\ is
\begin{subeqnarray}\label{linRN}
  \dot{\bar r}&=&N\;,\\
  \dot N&=&-N+2\bar r\;,\\
  \dot W&=&U\;,\\
  \dot U&=&-W-U\;,
\end{subeqnarray}
with the solution
\begin{subeqnarray}\label{solRN}
   W&=&C_1e^{-\frac{\tau}{2}}\sin(\frac{\sqrt{3}}{2}\tau+\theta)\;,\\
   U&=&C_1e^{-\frac{\tau}{2}}\sin(\frac{\sqrt{3}}{2}\tau+\frac{2\pi}{3}+\theta)\;.\\
   \bar r&=&C_2e^{\tau}-C_3\frac{1}{2}e^{-2\tau}\;,\\
   N&=&C_2e^{\tau}+C_3e^{-2\tau}\;.
\end{subeqnarray}
The oscillating solutions running into the f.p.\ for $\tau\to\infty$ require
$C_2=0$. For these solutions the term $C_3e^{-2\tau}$ does not 
describe the actual asymptotic behaviour of $\bar r$ and $N$, because   
the true behaviour is determined by nonlinear terms involving $W$ and
$U$. Integrating the resulting inhomogeneous linear equations for $\bar r$
and $N$ one finds (neglecting the non-leading term $C_3e^{-2\tau}$)
\begin{subeqnarray}\label{Nasy}
\bar r&=& C_2e^\tau+C_1^2e^{-\tau}\Bigl(-\frac{9}{28}\cos(\sqrt{3}\tau+2\theta)
      +\frac{\sqrt{3}}{28}\sin(\sqrt{3}\tau+2\theta)\Bigr)\;,\\
N&=&C_2e^\tau+C_1^2e^{-\tau}\Bigl(\frac{3}{7}\cos(\sqrt{3}\tau+2\theta)
      +\frac{2\sqrt{3}}{7}\sin(\sqrt{3}\tau+2\theta)\Bigr)\;.
\end{subeqnarray}
In order to run into the f.p.\ for $\tau\to -\infty$ it is necessary that
$C_1$ and $C_3$ vanish. 
From the form of Eqs.~(\ref{taueq}) one sees that the vanishing of 
$C_1$ implies $W\equiv 0$. Since the `exterior' ($r>1$) part of the limiting 
solution, when the number of nodes goes to infinity, 
runs into the RN f.p.\ for $\tau\to -\infty$ it must correspond to 
$W\equiv 0$.

\section{Classification}\label{chapclass}

The classification of solutions with a regular  origin proceeds
very similar to the one for the EYM system in \cite{BFM}. We
distinguish three different classes. The first one, which we call
singular (${\bf Sing}$) are solutions becoming singular according to
Eqs.~(\ref{sing}) with $r\to 0$ for some finite value $\tau_0$. This class may
be further subdivided into the classes ${\bf Sing}_n$ $(n=0,\ldots)$
containing singular solutions with $W_0^2>1$ and $W$ having $n$
nodes and the class ${\bf Sing}_\infty$ for singular solutions with
$W_0^2<1$. The second class are the globally regular solutions
(${\bf Reg}$) reaching the f.p.\ with $r=\infty$ described in
Eqs.~(\ref{rinf}). Again the class ${\bf Reg}$ can be subdivided 
according to the number of zeros of $W$.  
Finally there is a third class, the oscillating
solutions (${\bf Osc}$), running into the RN f.p. with $r=1$. Our aim is to
prove that any solution of Eqs.~(\ref{taueq}) with a regular origin belongs
to one of the classes ${\bf Sing}$, ${\bf Reg}$ or ${\bf Osc}$. This is made plausible by the
observation that the dilaton $\phi$ is a monotonically increasing 
function of $\tau$. 
Thus it can either stay bounded or diverge to 
$+\infty$; the latter can either happen for some finite $\tau_0$ or for 
$\tau\to\infty$. The difficulty is
to show that these alternatives lead precisely to the singular points 
described in Eqs.(\ref{sing}), (\ref{rinf}) and (\ref{solRN}) corresponding
to the three classes introduced above. 
The essential part of this claim are the equivalences stated in
Props. (1) to (3), whose proof will be based on a series of Lemmas 
(some of which can be found in the Appendix).  
In what follows we always consider solutions of Eqs.~(\ref{taueq}) starting
from a regular origin with $W=1$. 
\begin{Lemma}{}\label{gen}
\begin{itemize}

\item[i)]
The function $W$ can have neither maxima if\/ $W>1$ or
$0>W>-1$ nor minima if\/ $W<-1$ or $0<W<1$.

\item[ii)]
The functions $\dot\phi=1-N$, $f$ and $g$ are bounded below and 
if any of them is positive for some $\tau_0$ then it stays 
positive for all $\tau>\tau_0$.
In particular they are non-negative for solutions with a regular origin.

\item[iii)]
For finite $\tau$ the functions $U$ and $W$ are finite as long as 
$N$ is finite.

\end{itemize}
\end{Lemma}
\begin{Prf}{}
i) and ii) are trivial consequences of Eqs.~(\ref{taueq}) and
(\ref{auxeq}) and the fact that $1-N$, $f$ and $g$ are positive 
close to the origin, if they are regular at $r=0$. 

In order to prove iii) we remark that $\ln r$ and $U$ obey linear equations
and stay obviously finite as long as $W$ and $N$ are finite. 
Suppose now that $N$ is bounded and $W$ is unbounded,
then because of property i) $|W|$ actually tends to $\infty$. 
Following closely an argument put forward in \cite{BFM} Prop. 5 one finds that 
$W$ diverges for some finite $\tau_0$ like 
$W=\pm \sqrt 2/(\tau_0-\tau)+O(1)$. Plugging this into Eq.~(\ref{taueq}c)
leads to a divergence of $N$ contrary to our assumption.
\end{Prf}
\begin{Lemma}{}\label{Wlim}
Suppose $N\to-\infty$ at $\tau_0$ then $W$ and $\kappa \equiv r(1-N)$ 
have finite limits $W_0$
and $\kappa_0$ at $\tau_0$ with $W_0\neq\pm1$ and $\kappa_0>0$
(and as a consequence $r(\tau_0)=0$).
\end{Lemma}
\begin{Prf}{}
First we want to show that $W$ is bounded. Assume the contrary and without 
restriction $W>1$ and $U\geq 0$. 
If we can show that $U/W$ is bounded we get that also 
$\dot W/W=rU/W$ is bounded and thus $\ln W$ is bounded. 
Putting $z=W(1-N)-UT$ one gets
\begin{equation}
\dot z=r(1-N)U-z\geq -z\;,
\end{equation}
This implies that $z$ is bounded below and therefore since $T\to+\infty$ we
may assume$-z/T<1$ and estimate
\begin{equation}
\frac{d}{d\tau}\Bigl(\frac{U}{W}\Bigr)=T+(N-1)\frac{U}{W}-r\frac{U^2}{W^2}
<T\Bigl(1-\frac{U^2}{W^2}-\frac{z}{WT}\frac{U}{W}\Bigr)\;.
\end{equation}
The r.h.s.\ is negative, if $U/W>2$ and thus $U/W$ is bounded implying
the boundedness of $W$. Then also $U$ and $\dot W$ are bounded
implying the finiteness of $W(\tau_0)$. 

Next we want to show that $\kappa$ has a finite positive limit.
Putting $\bar f\equiv r^2f$ we get from Eq.~(\ref{auxeq}) 
\begin{equation}\label{rneq}
\dot{\bar f}=2(N-1)\bar f+4\dot{W}^2\;.
\end{equation}
Since $\int^{\tau_0}Nd\tau<0$ and $\dot{W}^2$ is integrable we get from 
Lemma \ref{linear} that 
$\bar f$ is bounded and hence also $\kappa$. The latter implies $r\to 0$
and thus $\int^{\tau_0}Nd\tau=-\infty$. Using once more Lemma \ref{linear}
we get $\bar f\to 0$. Since the boundedness of $U=\dot W/r$ yields 
$\dot W\to 0$ we get $\kappa^2\to (W_0^2-1)^2$.

In order to show that $\kappa$ stays away from zero we use again $g\geq 0$ 
to estimate
\begin{equation}
\dot{\kappa}=(N-1)\kappa+2rU^2+rT^2\geq -\kappa+4rU^2\geq -\kappa».
\end{equation}
This inequality shows that $\ln\kappa$ is bounded below
and consequently $W_0^2\neq\pm 1$. 
\end{Prf}
\begin{Lemma}{}\label{Nneg}
If $N<-1$ at some point $\tau_1$ then $N$ tends to $-\infty$  
for some finite $\tau_0>\tau_1$.
\end{Lemma}
\begin{Prf}{}
From $g\geq0$ we get $T^2\geq 2U^2-N(1-N)$ and from
Eq.~(\ref{taueq}c)
\begin{equation}\label{ineqN}
\dot N(\tau)\leq 1-N-4U^2+N-N^2\leq 1-N^2<0 \quad {\rm for} \quad 
   \tau\geq\tau_1
\end{equation}
and thus $N$ stays below $-1$. Suppose $N(\tau)<-1-\epsilon$ for
$\tau>\tau_1$ with
$\epsilon>0$ then $\dot N<-2\epsilon$. Hence $N$ is unbounded from below.
Using Lemma~\ref{Riccatilin} we get $N\to -\infty$ for some finite $\tau_0$.  

%
%
\end{Prf}

\begin{Lemma}{}\label{Wgreat}
If $W^2>1$ and  $WU\geq 0$ at some point then the
function $N$ tends to $-\infty$ for some finite $\tau_0$.
\end{Lemma}
\begin{Prf}{}
The proof follows essentially the one given in \cite{BFM}.
From  
\begin{equation}\label{Teq}
\dot T=2WU-NT, \quad \dot U=WT+(N-1)U\;,
\end{equation}
we get
\begin{equation}\label{TUineq}
\frac{d}{d\tau}\ln{|TU|}\geq 2\sqrt 2|W|-1\geq 2\sqrt 2-1\;,
\end{equation}
It follows that for any $c>0$ we can find some $\tau_1$ such that 
$2\sqrt{2}|TU|\geq c$ for $\tau>\tau_1$ and hence $T^2+2U^2\geq c$.
Eq.~(\ref{taueq}c) then yields $\dot N\leq 1-c-N$ implyig that $N$ eventually
becomes arbitrarily negative for large enough $\tau$ and thus 
Lemma \ref{Nneg} may be applied.
\end{Prf}
\begin{Lemma}{}\label{Rbig}
$r$ cannot have a maximum with $r_{\rm max}>\sqrt2$ for $W^2\leq 1$.
\end{Lemma}
\begin{Prf}{}
At a maximum of $r$ we have $N=0$ and $\dot N\leq 0$. On the other hand 
$g\geq 0$ implies $T^2\geq 2U^2$ for $N=0$ and thus
\begin{equation}\label{Rmax}
\dot N\geq 1-2T^2>1-\frac{2}{r^2}>0\;.
\end{equation}
\end{Prf}
\begin{Lemma}{}\label{Rbdd}
If $\ln r$ is bounded for
$\tau\to\infty$ then $N\to 0$.
\end{Lemma}
\begin{Prf}{}
Lemmas \ref{gen} and \ref{Nneg} imply $|N|\leq 1$. Suppose $\limsup N>0$,
then there exist some $\epsilon>0$ and a sequence of points $\tau_i\to\infty$
with $N(\tau_i)>\epsilon$. 
Since $\dot N\leq 2$ we get
$N(\tau)>\epsilon/2$ for $\tau_i-\epsilon/4<\tau<\tau_i$, implying
$\int^\infty N^2 d\tau=\infty$. Similarly the assumption $\liminf N<0$
leads to the divergence of $\int^\infty N^2 d\tau$.
On the other hand we have $0\leq g\leq (1-N)^2<4$
and thus
\begin{equation}\label{intg}
\infty>\int^\infty(\dot g-\dot N)d\tau=2\int^\infty(N^2+2U^2-N)d\tau\;.
\end{equation}
Since $\int^\infty Nd\tau=\ln r+$ const. is bounded this implies the boundedness 
of $\int^\infty N^2d\tau$ contradicting the assumptions $\limsup N>0$
and $\liminf N<0$ and hence $\lim N=0$.
 
\end{Prf}
\begin{Lemma}{}\label{Rinfty}
$r\to\infty$ for
$\tau\to\infty$ implies $N\to 1$ for $\tau\to\infty$.
\end{Lemma}
\begin{Prf}{}
According to Lemma~\ref{Wgreat} we may assume $W^2\leq 1$ and thus $T\to 0$
for $\tau\to\infty$. Thus for any $\epsilon>0$ there is some $\tau_\epsilon$
such that $|T|<\epsilon$ for $\tau>\tau_\epsilon$. From $g\geq 0$ we get
\begin{equation}
\frac{d}{d\tau}(1-N)=N-1+2U^2+T^2\leq 2T^2-(1-N)^2\leq 2\epsilon^2-(1-N)^2\;.
\end{equation}
Together with $1-N\leq 1$ implied by Lemma~\ref{Rbig} this shows that
$0<1-N<2\epsilon$ for $\tau>\tau_\epsilon+\frac{1}{2\epsilon^2}$.
Hence $N\to 1$ for $\tau\to\infty$.
\end{Prf}

\begin{Prop}{}\label{Rtozero}
The following are equivalent
\begin{itemize}
\item[i)] 
$\phi\to\infty$ for some finite $\tau_0$
\item[ii)] 
$r\to 0$ for some finite $\tau_0$
\item[iii)]
$N\to-\infty$ for some finite $\tau_0$
\item[iv)]
The solution belongs to ${\bf Sing}$
\end{itemize}
\end{Prop}
\begin{Prf}{}
Obviously iv) implies i) and i) implies ii).

ii) $\Rightarrow$ iii):
Since $\ln r$ diverges $N$ must be unbounded from below
and thus iii) follows from Lemma~\ref{Nneg}. 

iii) $\Rightarrow$ iv):
We have to show that $r\to 0$ and 
the functions $W,U,\kappa,\lambda$ and $\mu$ have a
finite limit at $\tau_0$. From Lemma~\ref{Wlim} we know that $r\to 0$, 
$W\to W_0\neq\pm 1$ and $\kappa\to\kappa_0>0$.
Thus we have only to prove that the r.h.s.\ of Eqs.~(\ref{Req}b,d,e) 
are integrable. For that reason we have to show that $U^2$ and $U^3$
are integrable, which will be achieved using arguments put forward in
\cite{BFM}.

The boundedness of $r$ and $1/\kappa$ implies that $|rN|^{-1}$ and $r^\epsilon$
are bounded for any $\epsilon>0$ and consequently
\begin{equation}
\int_\tau^{\tau_0} Nr^\epsilon d\tau=\frac{1}{\epsilon}r^\epsilon<\infty 
\end{equation}
and hence
\begin{equation}
\int_\tau^{\tau_0}\frac{1}{r^{1-\epsilon}} d\tau<\infty\;.
\end{equation}
Applying Lemma~\ref{linear} we get that 
$r^\epsilon U$ obeying the linear equation
\begin{equation}
(r^\epsilon U)^.=\frac{W(W^2-1)}{r^{1-\epsilon}}+(N-1+\epsilon)r^\epsilon U
\end{equation}
is bounded for $\epsilon>0$. 
As a consequence $|U^n|$ is integrable for any $n>0$.
According to Lemma~\ref{linear} this implies that $\mu$  
and consequently $\lambda$ and $U$ have a limit.
\end{Prf} 
\begin{Prop}{}\label{Rtoinf}
The following are equivalent
\begin{itemize}
\item[i)]
$\phi\to\phi_\infty<\infty$ for $\tau\to\infty$
\item[ii)]
$r\to\infty$ for $\tau\to\infty$
\item[iii)]
The solution belongs to ${\bf Reg}$
\end{itemize}
\end{Prop}
\begin{Prf}{}
Obviously iii) implies i) and i) implies ii).

ii) $\Rightarrow$ iii):
From Lemma~\ref{Rinfty} we know that $N\to 1$ for $\tau\to\infty$.
Thus $E$ is asymptotically monotonously increasing. Suppose $E\to\infty$, 
then $|\dot W|\to\infty$. Yet, this is not compatible with the boundedness
of $W$. Hence $\dot W$ must be bounded and in fact tend to zero for
$\tau\to\infty$ according to Lemma~\ref{Lyap}. Since $E$ has a limit also
$W$ has a limit, which must be a f.p.\ of Eqs.~(\ref{taueq}). The f.p.\ with
$W=0$ is, however, excluded, because $E$ is
asymptotically increasing and thus cannot tend to its infimum $-1$.
Therefore the solution belongs to ${\bf Reg}$. 

\end{Prf}
\begin{Prop}{}\label{Rone}
The following are equivalent
\begin{itemize}
\item[i)]
$\phi\to\infty$ for $\tau\to\infty$
\item[ii)]
$|\ln r|<c<\infty$ for all $\tau$
\item[iii)]
The solution belongs to ${\bf Osc}$
\end{itemize}
\end{Prop}
\begin{Prf}{}
Obviously iii) implies i).

i) $\Rightarrow$ ii):
From Lemma~\ref{Rbig} and Prop.~\ref{Rtoinf} we conclude that $r$ must be
bounded and thus $\int^\infty Nd\tau$ is bounded from above. 
Eq.~(\ref{intg}) also shows that $\int^\infty Nd\tau$ is bounded from below and thus $\ln r$
has also a positive lower bound.

ii) $\Rightarrow$ iii):
Lemma~\ref{Rbdd} implies $N\to 0$ for  $\tau\to\infty$. 
Now we can use Lemma~\ref{Lyap} to conclude 
that $E$ has a limit and $\dot W\to 0$. Thus also $W$ has a limit, which
must be a f.p.\ of Eqs.~(\ref{taueq}). Eq.~(\ref{taueq}c) is only compatible
with $N\to 0$, if the f.p.\ is $W=0$ and $r=1$.    
\end{Prf}
Putting together the Props.~\ref{Rtozero},~\ref{Rtoinf} and \ref{Rone} with
the possible behaviours of $\phi$ discussed at the beginning of this section
we obtain the `Classification Theorem'

\begin{Thm}{}\label{Class}  
Any solution of Eqs.~(\ref{taueq}) with a regular origin belongs to one of
the three classes ${\bf Sing}$, ${\bf Reg}$ or ${\bf Osc}$.
\end{Thm}

\section{Topology of `Moduli Space' and Existence Theorem}\label{chapex}

The method to prove the existence of at least one globally regular solution
for each number of nodes and a corresponding limiting solution with 
infinitely many
nodes used in \cite{BFM} can be almost literally translated to the 
present case.\footnote{The numerical analysis yields exactly one regular
solution for each node number and correspondingly one single limiting
solution}  
The proof is based on an analysis of the phase space as a function of
the parameter $b$ determining the solutions with a regular origin.
While the generic singular solutions (i.e.\ solutions in ${\bf Sing}$) 
correspond 
to open intervals of $b$ space, the $b$ values for regular solutions 
are isolated points accumulating at the value(s) for the limiting
solution(s).
As a first step we will study, what happens for very small and very large
values of $b$. The result is the same as for the EYM system. 
\subsection{Small b}\label{small}
\begin{Prop}{}\label{smallb}
If $b\neq0$ is small enough the solution with $W|_{r=0}=1$ belongs 
to the singular class ${\bf Sing}_0$ for $b<0$ or to ${\bf Sing}_1$ for
$b>0$ .   
\end{Prop}
{\bf Remark}: In view of Lemma~\ref{gen} the restriction for $b$ to be small
is unnecessary for $b<0$.
\begin{Prf}{}
The proof runs completely along the lines of the one in \cite{BFM}.
Rescaling $r\rightarrow |b|^{-\frac{1}{2}}r$ and $U\rightarrow
|b|^{-\frac{1}{2}}U$ we obtain from Eq.~(\ref{taueq}c)
\begin{equation}\label{scaled}
  \dot N =1-N-|b|(2U^2+T^2)\;
\end{equation}
and the $b$ independent boundary condition 
$\lim_{\tau\to-\infty}\frac{U}{r}=\mp2$. For $b=0$ the solution is 
$N\equiv 1$ and thus $r=e^\tau$. As was shown the resulting solution $W$ of the
pure YM system diverges like $W\sim \pm\sqrt{2}/(\bar\tau-\tau)$ 
for some finite $\bar\tau$ resp.\ $\bar r_\pm$ depending on the sign of $b$. 
The values of $\bar r_\pm$ have been determined in \cite{BFM} numerically
as $\bar r_+=\approx 5.317$ and $\bar r_-\approx 1.746$.
For small $b$ we obtain a
small perturbation of this solution as long as 
$|b|\int^\tau_{-\infty}(2U^2+T^2)d\tau'\ll1$. This condition still holds for 
$|W|\gg1$ and $|b|^{-\frac{1}{2}}r\gg1$, if $b$ is small.
Thus the Prop.\ follows from Lemma~\ref{Wgreat}.   
\end{Prf} 
\begin{figure}[p]
\hbox to\linewidth{\hss
        \epsfig{bbllx=39bp,bblly=159bp,bburx=577bp,bbury=572bp,%
        file=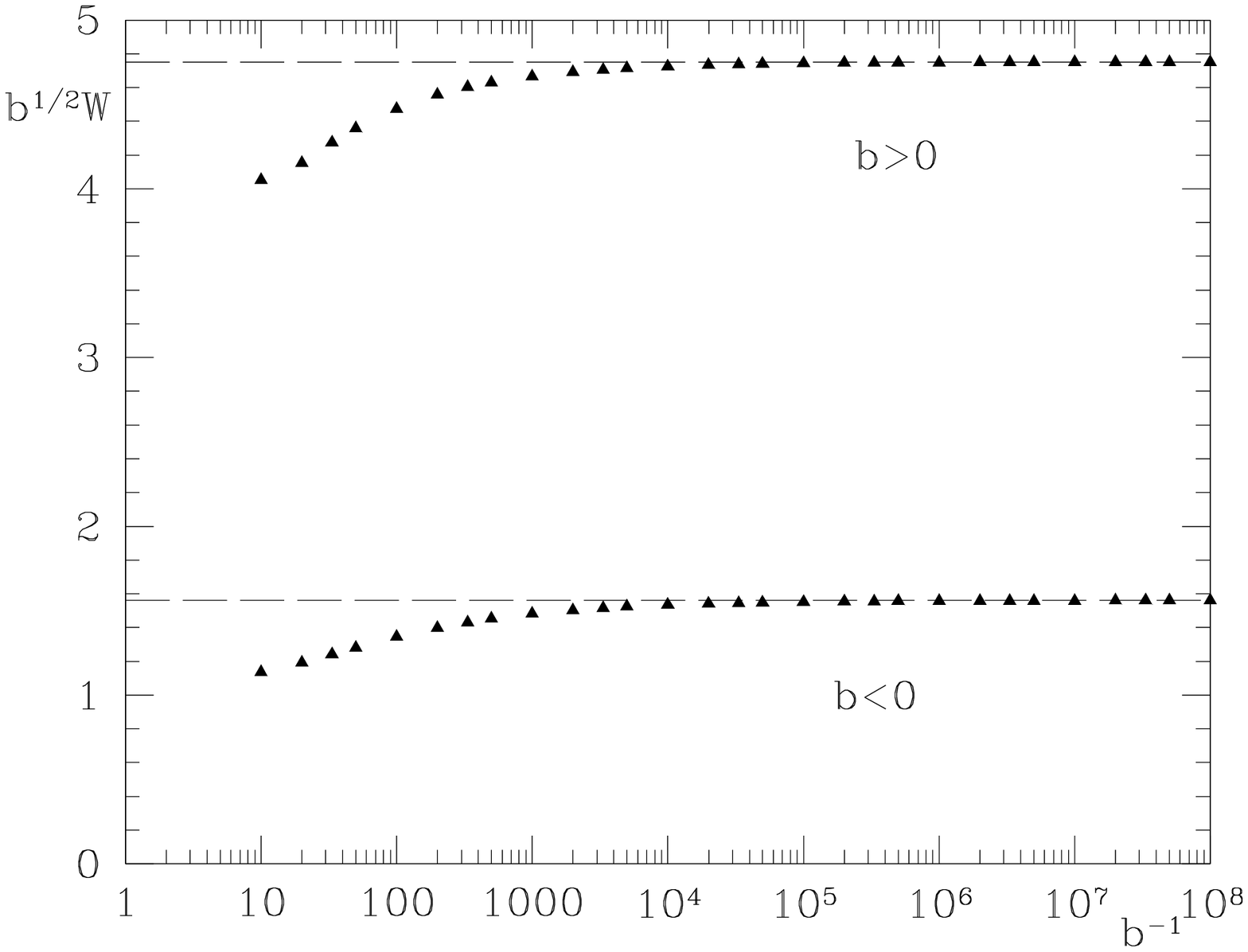,width=0.8\linewidth}\hss}
\caption[figsmall]{\label{figsmall}%
Numerical data for $W_0(b)$; the dashed lines represent the values of
Equ.~(\ref{asybs}) for the two signs of $b$ }
\end{figure}
Before we proceed to the case of large values of $b$ we shall derive the
asymptotic behaviour of $W_0$ as $r$ runs back to zero, similar to the one
given in \cite{BLM}. This proceeds in several steps. First we integrate the
Eq.~(\ref{taueq}c) for $N$ from $N=1$ to $N=0$ using
the pure YM solution becoming singular at $\tau=\bar\tau$. We obtain some 
value $\tau_e<\bar\tau$ from 
\begin{equation}\label{taue}
1=\frac{8|b|}{\bar r_\pm^2}\int_{-\infty}^{\tau_e}\frac{d\tau'}{(\bar\tau-\tau')^4}
=\frac{8|b|}{3\bar r_\pm^2(\bar\tau-\tau_e)^3}\;.
\end{equation}
This gives 
$|W_e|\approx\frac{\sqrt{2}}{(\bar\tau-\tau_e)}
\approx 3^{1\over3}\bar r_\pm^{2\over3}|b|^{-{1\over3}}/2$.
Here we have neglected the term $1-N$ in Eq.~(\ref{taueq}c) and pulled $r$ 
out of the integral; this yields an error of $O(1)$ to $W_e$, small by one
order in $|b|^\frac{1}{3}$. Correspondingly we get $r_e=\bar r_\pm$. 

The next step is to integrate Eqs.~(\ref{taueq}) from $N=0$ to $N=-\infty$
and to determine the value of $W_0$.
Since $rN$ is $O(1)$ for $N\to -\infty$ and $r_e=O(|b|^{-{1\over2}})$ we
rescale $\tau\to\tau |b|^{1\over2}$ and similarly the dependent variables 
$W\to W^|b|^{-{1\over2}}$ etc. 
Keeping only leading terms as $b\to 0$ we obtain from
Eqs.~(\ref{taueq})
\begin{subeqnarray}\label{bsmall}
  \dot r&=&rN\;,\\
  \dot W&=&rU\;,\\
  \dot N &=&-2U^2-T^2\;,\\
  \dot U&=&WT+NU\;,\\
  \dot T&=&2WU-NT\;
\end{subeqnarray}  
with the constraint $rT-W^2=0$. Due to the scaling the boundary conditions at 
$\tau_e$ are changed to $\tau\to-\infty$ and $r=\bar r_\pm, W=U=T=N=0$.
In order to perform the limit $r\to 0$ we
introduce the analogue of the variables used in Eq.~(\ref{Req}) putting
$\kappa=-rN$, $\lambda=WT+NU$ and $f=N^2+2U^2-T^2$. The Eqs.~(\ref{bsmall})
imply $\dot f=0$ and hence $f\equiv 0$. Using the new variables we
obtain
\begin{subeqnarray}\label{bsma}
  \dot r&=&-\kappa\;,\\
  \dot W&=&rU\;,\\
  \dot U&=&\lambda\;,\\
  \dot \lambda&=&(3W^2-4U^2)U\;\\
  \dot\kappa&=&4rU^2\;.
\end{subeqnarray}    
These equations have to be integrated from the highly degenerate f.p.\
$W=U=\kappa=\lambda=0, r=\bar r_\pm$ attained at $\tau=-\infty$.
In order to lift the degeneracy we use a `blow up' in the direction $W$ 
\cite{Dumortier} introducing 
\begin{equation}\label{blowup}
\bar U=\frac{U}{W^2},\quad \bar\kappa=\frac{\kappa}{W^3}\quad {\rm and}
\quad \bar\lambda=\frac{\lambda}{W^3}
\end{equation}
and using $\ln{|W|}$ as new independent variable. Thus we obtain the equations
\begin{subeqnarray}\label{wequ}
  W\frac{dr}{dW}&=&-\frac{W^2 \bar\kappa}{r\bar U}\;,\\
  W\frac{d\bar U}{dW}&=&\frac{\bar\lambda}{r\bar U}-2\bar U\;,\\
  W\frac{d\bar\lambda}{dW}&=&\frac{3-3r\bar\lambda-4W^2\bar U^2}{r}\;\\
  W\frac{d\bar\kappa}{dW}&=&-(4\bar U+3\bar\kappa)\;.
\end{subeqnarray}    
\begin{figure}[p]
\hbox to\linewidth{\hss
        \epsfig{bbllx=37bp,bblly=173bp,bburx=573bp,bbury=565bp,%
        file=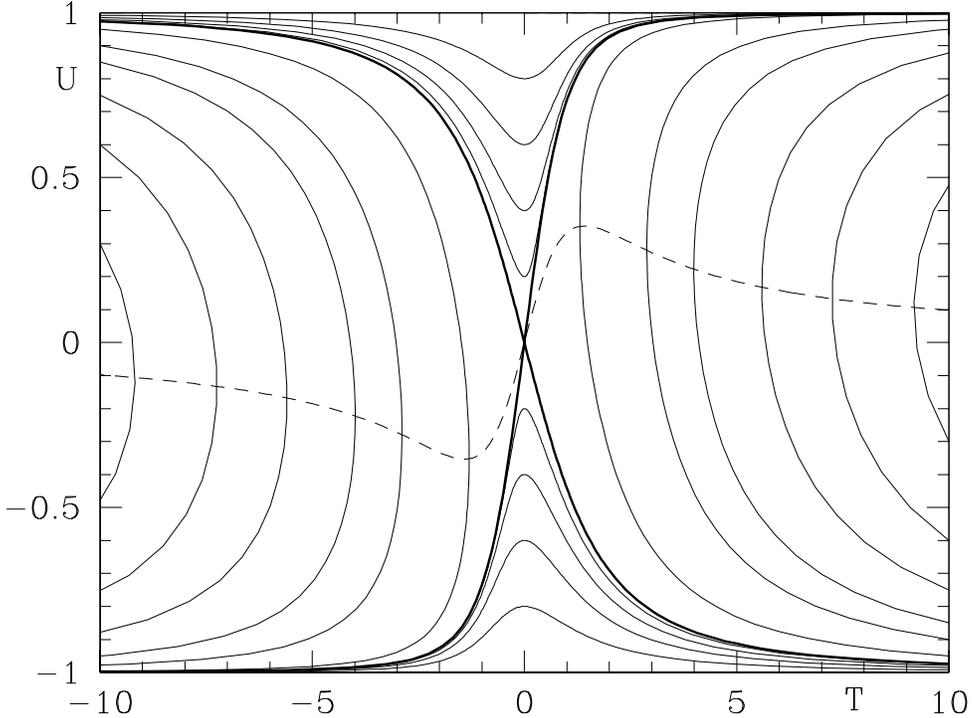,width=0.8\linewidth}\hss}
\caption[figph]{\label{figph}%
The phase space of Eqs.~(\ref{simple}); on the dashed curve 
$U=\frac{T}{2+T^2}$ the orbits are vertical}
\end{figure}

The f.p.\ has now moved to the point 
\begin{equation}\label{blow}
W=0,\quad r=\bar r_\pm,\quad\bar U=\pm\frac{1}{\sqrt{2}\bar r_\pm},\quad
\bar\kappa=\mp\frac{2\sqrt{2}}{3\bar r_\pm}
\quad {\rm and}
\quad \bar\lambda=\frac{1}{\bar r_\pm}\;.
\end{equation}
Linearisation at this f.p.\ yields exclusively negative eigenvalues
and thus the degeneracy has been removed. At the same time this shows that
there are no adjustable parameters at the f.p.\ as to be expected.
 
Numerical integration of Eqs.~(\ref{wequ}) from the f.p.\ to $r=0$
results in a finite value of $W$ proportional to $\bar r_\pm$. Taking into
account the scaling factor $|b|^{-\frac{1}{2}}$ we find
\begin{equation}\label{asybs}
W_0\approx 0.89369|b|^{-\frac{1}{2}}\bar r_\pm\;.
\end{equation}
Fig.~\ref{figsmall} shows a plot of numerically obtained data for $W_0$;
the dashed lines represent the values of Eq.~(\ref{asybs}) for
the two different signs of $b$.

Next we turn to solutions with large values of $b$. Again the situation
resembles very much the EYM case.
\subsection{Large b} 
\begin{Prop}{}\label{largeb}
If $b\gg0$ is large enough the solution with $W|_{r=0}=1$ belongs
to the singular class ${\bf Sing}_\infty$.
\end{Prop}
\begin{Prf}{}
We put $r=\bar r/b$ and $W=1+\bar W/b$. Keeping only leading terms
we obtain from Eqs.~(\ref{taueq})
\begin{subeqnarray}\label{blarge}
  \dot{\bar r}&=&N\bar r;,\\
  \dot{\bar W}&=&\bar r U\;,\\
  \dot T&=&2U-NT\;,\\
  \dot U&=&T+(N-1)U\;,\\
  \dot N&=&1-N-2U^2-T^2\;
\end{subeqnarray}    
with the constraint $2\bar W-\bar r T=0$ and the $b$ independent boundary
condition $\lim_{\tau\to-\infty}\frac{U}{\bar r}=\mp2$.
The combination $z=N-1+TU$ obeys the simple equation $\dot z=-z$. In view of
the initial condition $\lim_{\tau\to-\infty}z=0$ we get $z\equiv 0$.
This allows us to remove $N$ from the $T,U$ system to obtain 
\begin{subeqnarray}\label{simple}
  \dot T&=&U(2+T^2)-T\;,\\
  \dot U&=&T(1-U^2)\;.
\end{subeqnarray}    
with the boundary condition $\lim_{\tau\to-\infty}U=\lim_{\tau\to-\infty}T=0$
such that $\lim_{\tau\to-\infty}\frac{U}{T}=1$.
It is straightforward to analyze the flow of this 2d system 
(Fig.~\ref{figph} shows a plot).  
The point 
$U=T=0$ is a hyperbolic f.p.\ with the eigenvalues $-2$ and $1$. No orbits can
cross the lines $U=\pm1$. The solution
with the relevant boundary condition from above corresponds to the
separatrix for the eigenvalue $1$. 
For $1>|U|>|T|/(2+T^2)$ the orbits are monotonously approaching one of 
the lines
$U=\pm1$ for $|T|\to\infty$. This shows that eventually $N<-1$ even 
taking into account the correction terms of $O(1/b)$ in Eqs.~(\ref{taueq}).
Applying Lemma~\ref{Nneg} together with $W=1-O(1/b)$ proves the proposition.     
\end{Prf}
\subsection{Existence Theorem}

As for the EYM theory one can characterize the neighbourhood of the sets 
${\bf Reg}_n$ and ${\bf Osc}$.
\begin{Prop}{}\label{regn}
Given $b_n\in {\bf Reg}_n$ for any $n$ then all $b\neq b_n$ 
sufficiently close to $b_n$ are either in ${\bf Sing}_n$ or 
${\bf Sing}_{n+1}$.
\end{Prop}
\begin{Prop}{}\label{osc}
Given $b_\infty\in {\bf Osc}$ and some $n_0$ then all $b\neq b_\infty$ 
sufficiently close to $b_\infty$ are either in ${\bf Sing}_\infty$ or in 
$\bigcup_{n\geq n_0}({\bf Reg}_n\cup {\bf Sing}_n)$.
\end{Prop}  

Equipped with the knowledge, what happens for large and small values of $b$
one proves
\begin{Thm}{}\label{exist}
\begin{itemize}
\item [i)]
The sets ${\bf Reg}_n$ and ${\bf Osc}$ are all nonempty, i.e., for each
$n=0,1,2,\ldots$ there exists a globally regular solution with $n$ zeros
of $W$ for at least one $b_n\in{\bf Reg}_n$ and there exists an
oscillating solution with $N>0$ for all $\tau$ and $r\to1$ for 
$\tau\to\infty$ for at least one $b_\infty\in{\bf Osc}$.

\item [ii)]
The union $\bigcup\limits_{n\ge0}{\bf Reg}_n$ has accumulation points
that are contained in ${\bf Osc}$, i.e., there exists at least
one sequence of globally regular solutions and one oscillating solution
$W_\infty$ such that $W_n(r)\to W_\infty(r)$ for $r<1$ and $W_n(r)\to0$
for $r\ge1$ for $n\to\infty$.
\end{itemize}
\end{Thm}

The proofs of the Props.~\ref{regn} and \ref{osc} and of the Theorem 
can be literally taken from \cite{BFM} and will not be repeated here.

{\bf Remark:} The existence of at least one regular solution for any $n$
was already proven in \cite{Hastings}. As mentioned above, the numerical
results show that there is exactly one regular solution for any $n$ and 
correspondingly only one limiting solution. As in the EYM case there is
however no uniqueness proof available or in view.

\section{Scaling law for large n}\label{chapstab}
In \cite{BFM} a remarkably well satisfied asymptotic scaling law for the
parameters of the regular solutions with a large number $n$ of nodes of $W$
was formulated. The derivation was based on the observation that for
solutions with many nodes three distinctive regions could be observed.
In an inner region the solutions are well approximated by the limiting
oscillating solution with infinitely many zeros. 
This region extends between the origin
and $r=1$. Furthermore there is an asymptotic region for $r>r_n\gg0$, where the
solutions are close to the flat solution connecting the f.p.s with 
$W=\pm1$, $U=0$ and $W=U=0$. In the intermediate region extending between
$r=1$ and $r=r_n$ the functions $W$ and $U$ stay small and thus
the equations for $W$ and $U$ can be linearized on the metric background
given by the extremal Reissner-Nordstr{\o}m solution. Boundary conditions
for these linearized YM equations are obtained by matching with the
solutions obtained in the inner and outer regions.
The same type of scaling law can be obtained here through more or less 
identical reasoning.

On region I defined by $0\leq r\leq 1$ the solutions are approximated by the
limiting solution running into the f.p.\ $W=U=N=0$. The corresponding
behaviour near $r=1$ is given by Eq.~(\ref{solRN}) neglecting the $C_3$
term. The ($n$ dependent) coefficient $C_2$ has to be positive to allow the
solution to reach region II with $r>1$. By a suitable shift in $\tau$ we can
always achieve $C_1=1$. The phase $\theta$ is adjusted such that
$\frac{\sqrt{3}}{2}\tau+\theta=m\pi$ at the $m^{\rm th}$ zero of $W$.

In region II the Eqs.~(\ref{taueq}b,d) for $W$ and $U$ are linearized
in the background of the solution Eq.(\ref{eRN}).        
Surprisingly the solution is identical to that of \cite{BFM}
\begin{subeqnarray}\label{regII}
  r_{\rm II}(\tau)&=&1+C_{2,n}e^\tau;,\\
  N_{\rm II}(\tau)&=&\frac{C_{2,n}e^\tau}{1+C_{2,n}e^\tau}\;,\\   
  W_{\rm II}(\tau)&=&
  e^{-{1\over2}\tau}\sin({\sqrt{3}\over2}\tau+\theta)
  +C_{2,n} e^{{1\over2}\tau}
  \sin({\sqrt{3}\over2}\tau+{\pi\over3}+\theta)\;.
\end{subeqnarray}
In region III, where we have $r\gg1$ and $1-N\ll1$, we take the flat solution 
connecting the f.p.s with $W=\pm1$, $U=0$ and $W=U=0$. In the region where $W$
is small this solution can be approximated by
\begin{subeqnarray}\label{regIII}
r_{\rm III}(\hat\tau)&=&c_ne^{\hat\tau}\;,\\
W_{\rm III}(\hat\tau)&=&
  \pm\hat C_{1,n}e^{{1\over2}\hat\tau}
  \sin({\sqrt{3}\over2}\hat\tau+{\pi\over3}+\hat\theta)\;,
\end{subeqnarray}
with the normalization $W_{\rm III}\to\pm(1-e^{-\hat\tau})$.
Again the phase $\hat\theta$ is adjusted such that

${\sqrt{3}\over2}\hat\tau+{\pi\over3}+\hat\theta=-m\pi$ at
the last but $m^{\rm th}$ zero of $W$. 
Matching $r_{\rm II},W_{\rm II}$ with $r_{\rm III},W_{\rm III}$ we obtain
\begin{subeqnarray}\label{match}
C_{2,n}e^{\tau}&=&c_n e^{\hat\tau}\;,\\
C_{2,n}e^{{1\over2}\tau}&=&\hat C_1e^{{1\over2}\hat\tau}\;,\\            
{\sqrt{3}\over2}\tau+\theta&=&{\sqrt{3}\over2}\hat\tau+\hat\theta+n\pi\;, 
\end{subeqnarray}
where $n$ is the total number of zeros of $W$.
Eliminating $\tau$ and $\hat\tau$ we obtain 
\begin{equation}\label{quant1}
C_{2,n}=\hat C_1e^{{1\over\sqrt{3}}(\theta-\hat\theta-n\pi)}
  \equiv C_{2,0}e^{-n{\pi\over\sqrt{3}}}
\end{equation}
and 
\begin{equation}\label{quant2}
c_n=C_{2,n}^{-1}\hat C_1^2\equiv c_0 e^{n{\pi\over\sqrt{3}}}\;.
\end{equation}
Since the coefficient $C_2(b)$ has to vanish for the limiting solution,
i.e.\ $b=b_\infty$, we get 
\begin{equation}\label{vanish}
C_2(b)=\frac{\partial C_2}{\partial b}(b_\infty)(b-b_\infty)
         +O((b-b_\infty)^2)\;.
\end{equation}
Numerical integration of the limiting solution and its
variation with respect to $b$ yields  
\begin{equation}\label{limitsl}
\theta\approx1.562209\qquad{\rm and}\qquad
   C_2\approx0.835060\cdot(b_\infty-b)\;,
\end{equation}
while numerical integration of the flat YM equations yields \cite{BFM}
\begin{equation}\label{flat}
\hat\theta\approx0.339811\qquad{\rm and}\qquad
  \hat C_1\approx0.432478\;.
\end{equation}
From Eqs.~(\ref{quant1},\ref{quant2}) we obtain 
\begin{equation}\label{largen}
b_n=b_\infty-1.04894\cdot e^{-n\alpha}\,,\quad
{\rm and}\quad  c_n=0.213530\cdot e^{n\alpha}\,,
\end{equation}
with $\alpha={\pi\over\sqrt3}\approx1.81380$ and
$e^\alpha\approx6.13371$.

\begin{table}
\caption[tabpar]{
Parameters $b$ and $c$ of regular solutions;numerical results versus
asymptotic formula Eq.~(\ref{largen})
\label{tabpar}}
\begin{center}\begin{tabular}{c|cc|cc}
 $n$&$b_{\rm num}$&$b_{\rm asy}$& $c_{\rm num}$&$c_{\rm asy}$\\
\hline\hline
 1& $0.2608301456037$&$0.20848171$&$7.525748e-01$&$1.309730e+00$\\\hline
 2& $0.3535180998051$&$0.35161335$&$7.320406e+00$&$8.033504e+00$\\\hline
 3& $0.3750018038731$&$0.37494861$&$4.852149e+01$&$4.927516e+01$\\\hline
 4& $0.3787544658699$&$0.37875304$&$3.014792e+02$&$3.022394e+02$\\\hline
 5& $0.3793733291287$&$0.37937329$&$1.853097e+03$&$1.853848e+03$\\\hline
 6& $0.3794744134274$&$0.37947441$&$1.137028e+04$&$1.137096e+04$\\\hline
 7& $0.3794908985808$&$0.37949090$&$6.974585e+04$&$6.974616e+04$\\\hline
 8& $0.3794935863472$&$0.37949359$&$4.278045e+05$&$4.278025e+05$\\\hline

\end{tabular}\end{center}

\end{table}
\begin{table} 
\caption[tabm]{
Masses of regular solutions;
numerical results versus asymptotic formula Eq.~(\ref{masses})
\label{tabm}}
\begin{center}\begin{tabular}{c|cc}
 $n$&$M_{\rm num}$&$M_{\rm asy}$\\
\hline\hline
 1& $0.80380777208$&$0.79578582$\\\hline
 2& $0.96559851724$&$0.96670624$\\\hline
 3& $0.99432009439$&$0.99457200$\\\hline
 4& $0.99907210998$&$0.99911505$\\\hline
 5& $0.99984867329$&$0.99985572$\\\hline
 6& $0.999975327358$&$0.99997648$\\\hline
 7& $0.9999959774969$&$0.99999617$\\\hline
 8& $0.99999934419598$&$0.999999374$\\\hline

\end{tabular}\end{center}

\end{table}
\begin{table} 
\caption[tabquot]{
Quotients of parameters of regular solutions
\label{tabquot}}
\begin{center}\begin{tabular}{c|c|c|c}
 $n$&$\Delta b_n$&$\Delta c_n$& $\Delta M_n$\\
\hline\hline
 1& $4.56821367$&$9.72714739$&$5.70301662$\\\hline
 2& $5.78233316$&$6.62825122$&$6.05669973$\\\hline
 3& $6.07360547$&$6.21331291$&$6.12131340$\\\hline
 4& $6.12385583$&$6.14668276$&$6.13170021$\\\hline
 5& $6.13210160$&$6.13582559$&$6.13338085$\\\hline
 6& $6.13345216$&$6.13404859$&$6.13365394$\\\hline
 7& $6.13370605$&$6.13376280$&$6.13369695$\\\hline

\end{tabular}\end{center}

\end{table}

In \cite{BFM} also the asymptotic formula for the mass
$M_n=1-\frac{3}{2}C_{2,n}$ was derived. The same formula is supported by our
numerical data in the present case, but in contrast to the EYM theory
we were not able to find a simple derivation. Putting in numbers yields
\begin{equation}\label{masses}
M_n=1-1.25259\cdot e^{-n\alpha}\,,
\end{equation}
Tables~\ref{tabpar} and \ref{tabm} contain a comparison of the numerically
determined parameters $b_n,c_n$ and $M_n$ with the asymptotic values
computed with the formulas from above. Table~\ref{tabquot} displays the
quotients $\Delta b_n=\frac{b_\infty-b_n}{b_\infty-b_{n+1}}$, 
$\Delta c_n=\frac{c_{n+1}}{c_n}$ and $\Delta M_n=\frac{1-M_n}{1-M_{n+1}}$
of the numerical data. 
All of them approach rapidly the value $e^{\pi/\sqrt 3}$.

\section*{Appendix:}
\begin{Lemma}{}\label{linear}
Consider a solution $y$ of the linear differential equation $\dot y=a+by$ in
some interval $\tau_0\le\tau<\tau_1$ with $|a|$ integrable. If
\begin{equation}
c(\tau',\tau)=\int_{\tau'}^\tau b(\tau'')d\tau''\;,
\end{equation}
is bounded from above for $\tau_0\le\tau'\le\tau<\tau_1$ then $y$ is
bounded; if $c(\tau',\tau)$ has a limit as $\tau\to\tau_1$ then $y(\tau)$
has a limit; if $c(\tau',\tau_1)=-\infty$ then $y(\tau_1)=0$.
\end{Lemma}
\begin{Prf}{}
All properties are implied by the explicit form
\begin{equation}
y(\tau)=y(\tau_0)e^{c(\tau_0,\tau)}
  +\int_{\tau_0}^\tau a(\tau')e^{c(\tau',\tau)}d\tau'\;.
\end{equation}
\end{Prf}
\begin{Lemma}{}\label{Riccatilin}
Suppose $y$ obeys the inequality $\dot y\leq a+by-y^2$. If $a$ is
bounded from above and $b$ is bounded for $\tau\ge\tau_0$, then $y$ is  
either bounded for all $\tau\ge\tau_0$ or diverges to $-\infty$ for some
finite $\tau_1>\tau_0$.
\end{Lemma}
\begin{Prf}{}
Let $A$, $B$ be positive constants such that $a<A$ and $|b|<B$. We can   
estimate $\dot y<0$ for $|y|>C=\sqrt{2A}+2B$ and therefore $y$ is bounded
from above. Furthermore $y$ monotonically decreases and $(1/y)\dot{}>1/2$
for $y<-C$, and thus $y\to-\infty$ for some finite $\tau_1$.
\end{Prf}

\begin{Lemma}{}\label{Lyap}
Suppose there is some open invariant subset $I$ of the phase space
of the system Eq.~(\ref{taueq}) such that $2N-1$ has a definite
sign in $I$. Then for any trajectory $(2N-1)r^2U^2$ vanishes on
all its limit points for $\tau\to\infty$ in $I$.
\end{Lemma}
\begin{Prf}{}
Eq.~(\ref{energy}) shows that the function $E$ is monotonous along
trajectories in $I$ and hence serves as a `Lyapunov Function'.
According to Lemma 11.1 of \cite{Hart} $\dot E=(2N-1)r^2U^2$
vanishes on the limit points of the solution.
\end{Prf}
\begin{Coro}{}\label{fixp}
Suppose $N\to 0$, $r\to r_0\neq 0$ for $\tau\to\infty$ and $W$ and
$U$ stay bounded, then the solution tends to a f.p.\ of
Eqs.~(\ref{taueq}) with $U=0$ and $W=0$ or $W^2=1$.
\end{Coro}
\begin{Prf}{}
From Lemma \ref{Lyap} we know that $\dot W\to 0$ for $\tau\to\infty$. 
Since $E$ is bounded it has a limit
for $\tau\to\infty$ and thus also $W$ has a limit, which must be a
f.p.\ of Eq.~(\ref{taueq}f) otherwise $\dot W$ would not tend to zero.
\end{Prf}

\section*{Acknowledgements}
I am indebted to P.~Breitenlohner for many discussions and for help with the
numerical computations.

\end{document}